\documentclass[preprint,12pt]{elsarticle}

\usepackage{amssymb}
\usepackage{amsmath}
\usepackage{tabularx}
\usepackage{booktabs}
\usepackage{siunitx}
\usepackage{float}
\usepackage{makecell}

\journal{Journal of Computational Physics}

\begin{document}
\begin{frontmatter}



\title{From Mesh to Neural Nets: A Multi-Method Evaluation of Physics-Informed Neural Networks and Galerkin Finite Element Method for Solving Nonlinear Convection-Reaction-Diffusion Equations}


\author[1]{Fardous Hasan\corref{cor1}}
\cortext[cor1]{Corresponding author}
\ead{Fardous.Hasan@uib.no}
\affiliation[1]{organization={Department of Clinical Dentistry, University of Bergen},
            city={Bergen},
            country={Norway}}
            
\author[2]{Hazrat Ali}
\ead{hazrat.ali@du.ac.bd}
\affiliation[2]{organization={Department of Mathematical Science, University of Texas},
            city={El Paso},
            country={USA}}

\author[3]{Hasan Asyari Arief}
\ead{hasv@norceresearch.no}
\affiliation[3]{organization={NORCE Norwegian Research Centre AS},
            addressline={},
            city={Bergen},
            country={Norway}}

\begin{abstract}
Non-linear convection-reaction-diffusion (CRD) partial differential equations (PDEs) are crucial for modeling complex phenomena in fields such as biology, ecology, population dynamics, physics, and engineering. Numerical approximation of these non-linear systems is essential due to  the challenges of obtaining exact solutions. Traditionally, the Galerkin finite element method (GFEM) has been the standard computational tool for solving these PDEs. With the advancements in machine learning, Physics-Informed Neural Network (PINN) has emerged as a promising alternative for approximating non-linear PDEs.

In this study, we compare the performance of PINN and GFEM by solving four distinct one-dimensional CRD problems with varying initial and boundary conditions and evaluate the performance of PINN over GFEM. This evaluation metrics includes error estimates, and visual representations of the solutions, supported by statistical methods such as the root mean squared error (RMSE), the standard deviation of error, the  the Wilcoxon Signed-Rank Test and the coefficient of variation (CV) test.

Our findings reveal that while both methods achieve solutions close to the analytical results, PINN demonstrate superior accuracy and efficiency. PINN achieved significantly lower RMSE values and smaller standard deviations for Burgers’ equation, Fisher’s equation, and Newell-Whitehead-Segel equation, indicating higher accuracy and greater consistency. While GFEM shows slightly better accuracy for the Burgers-Huxley equation, its performance was less consistent over time. In contrast, PINN exhibit more reliable and robust performance, highlighting their potential as a cutting-edge approach for solving non-linear PDEs.

\end{abstract}



\end{frontmatter}




\section{Introduction}\label{sec1}

Nonlinear Partial Differential Equations (PDEs) are a specific type of PDEs that introduce nonlinearity into the equations, making them significantly more complex and versatile. These equations are widely used in both physical and computational sciences to simulate behaviors such as shock wave propagation, turbulence, and other nonlinear dynamic phenomena \cite{wang2024approximating}. They are commonly applied in areas like fluid dynamics, plasma physics, solid mechanics, environmental science, and mathematical biology, where they help modeling the complex interactions and characteristic patterns. \cite{yang2023new, ghergu2011nonlinear, ahmad2022new}. However, the inherent complexity and non-linearity of convection-reaction-diffusion (CRD) PDEs present significant challenges in achieving accurate and efficient solutions. Accurately solving these equations is vital for predicting complex behaviors in various systems, such as fluid dynamics, quantum mechanics, and biological processes. Even small inaccuracies can lead to significant errors in simulations, which might compromise the reliability of research and practical applications.

The challenges of solving non-linear PDEs arise from their complexity and the diverse range of phenomena they represent. Traditional analytical methods often fall short, especially when dealing with complicated geometries, non-linear interactions, and varying boundary conditions. Consequently, advanced numerical techniques have been developed to address these issues.

These advanced numerical techniques include traditional methods such as the Finite Element Method (FEM), which is widely used for its flexibility in handling complex geometries and boundary conditions \cite{liu2015finite}. Another significant method in this category is the Galerkin Finite Element Method (GFEM). The GFEM is a type of weighted residual method that uses piecewise trial functions to approximate the solution over each finite element in the domain. It offers a powerful and flexible approach for the numerical solution of differential equations, enhancing the capabilities of the traditional Finite Element Method. In addition to these, the Finite Difference Method (FDM), the Finite Volume Method (FVM) and spectral element method (SEM) are also commonly employed, each with its own strengths in stability and accuracy \cite{iserles2009first, ciarlet1990handbook, patera1984spectral}. Recently, deep learning approaches have emerged as powerful alternatives to conventional techniques. Methods like Deep Galerkin Method (DGM), Physics-Informed Neural Network (PINN) and DeepONet use neural networks to approximate solutions of non-linear PDEs \cite{sirignano2018dgm, deng2022approximation, raissi2019physics}. By encoding the physics of the problem directly into the training process, these models can efficiently learn from data while satisfying the underlying equations, offering promising results in both accuracy and computational efficiency.

Among these methods, the GFEM has gained prominence for its robustness and flexibility in handling complex problems \cite{multiphysics2017finite}. GFEM effectively discretizes the domain and accommodates non-linearities, providing stable and accurate solutions. In recent years, the emergence of PINN has offered a revolutionary approach to solving PDEs. By integrating the principles of physics directly into the neural network training process, PINNs allow for the efficient approximation of solutions while inherently satisfying the governing equations and boundary conditions.

The GFEM can be computationally expensive, especially for large-scale problems requiring fine meshes and higher-order elements. Its implementation is complex and demands a thorough understanding of the underlying mathematics and numerical techniques. The accuracy of GFEM solutions depends heavily on mesh quality, with poor meshes leading to inaccuracies. Additionally, handling complex boundary conditions often necessitates special techniques, and the method may face convergence issues with highly nonlinear problems. Other drawbacks include limited software maturity and increased memory demands. Conversely, the PINN algorithm offers a mesh-free approach, transforming PDE solving into a loss function optimization problem, suitable for complex geometries and high-dimensional spaces. PINNs eliminate the need for labeled data and embed governing physics equations within a residual network \cite{cuomo2022scientific}. They generalize well, predicting solutions for different resolutions without retraining. However, PINNs require substantial training time and computational resources and can be sensitive to training data quality, potentially leading to overfitting. Studies show FEM outperforms PINN in accuracy and computation time, but evaluating a trained PINN is significantly faster than interpolating FEM results on a new mesh \cite{grossmann2024can, xu2023transfer}.

In this study, we conducted a comprehensive comparative analysis between the advanced deep learning-based PINN and the traditional GFEM. These methods were applied to solve four prominent nonlinear CRD PDEs, including Burgers’ Equation, Fisher’s Equation, Burgers-Huxley Equation, and the Newell-Whitehead-Segel Equation.
The objective of this study is to systematically evaluating the performance, accuracy, and consistency of PINN over GFEM across a range of non-linear PDEs.

This research introduces a novel multi-method evaluation framework that integrates tabular numerical comparisons, graphical visualizations, and rigorous statistical analyses, including Root Mean Squared Error (RMSE), standard deviation, the Wilcoxon Signed-Rank Test and the  Levene’s variability test, to validate the performance of the methods. This comprehensive approach provides a deeper understanding of the performance of PINNs and GFEM, ensuring robust and reliable results.

The structure of this article is as follows: Section \ref{sec2} reviews related works. Section \ref{sec3} details the methodologies of GFEM and PINN. Section \ref{sec4} describes the experimental settings, while Section \ref{sec5} presents and discusses the computational results for solving four nonlinear PDEs, comparing the performance of GFEM and PINN.

\section{Related Works}\label{sec2}

The study of non-linear PDEs has garnered significant attention, particularly in the context of convection-reaction-diffusion equations. Various methods, ranging from traditional numerical techniques to modern deep learning-based approaches, have been explored for their efficacy in solving these complex equations. For example, the work of Yang Liu et al. (2015) \cite{liu2015finite} showed that the Finite Element Method (FEM) is very effective for solving complex equations, including those with higher-order derivatives and nonlinear terms. Their method provides stable and accurate solutions, even for difficult time-dependent problems, ensuring precise results across the entire computational area.The study of Ahmed et al. (2022) \cite{ahmed2022finite}, demonstrated  the robustness of FVM in maintaining flux consistency across control volumes, which is crucial for preserving mass, momentum, and energy throughout the computational domain. The method's ability to handle complex geometries makes it particularly suitable for convection-dominated problems, ensuring the conservation laws are respected. Previously, the modified version of FVM, introduced by Xu in 2018 \cite{xu2018modified}, achieved an accuracy level nearly equivalent to the exact solution, without inducing any unphysical oscillations for convection/reaction-dominated problems. Since the main advantage of FEM and FVM is its stability, FEM is more mathematically robust approach than FVM \cite{lopes2021analysis}. Another study of Xu (2022) \cite{xu2022finite} further refined this approach by developing Finite Volume Scheme (FVS) that not only achieves significantly higher accuracy but also exhibits unparalleled stability compared to the traditional FVM based on central differencing. 

Another numerical technique, called FDM, has also been extensively studied. The work of Tsega in 2024 \cite{tsega2024numerical} highlighted the robustness of FDM in simulating advection-diffusion-reaction problems with variable coefficients. This method has proven effective in capturing the intricate dynamics of non-linear PDEs, making it a valuable tool in numerical analysis. Similarly, the work of Giraldo in 2023 \cite{giraldo2023adaptive}, highlighted the performance of another numerical technique called FEM in advection-dominated problems. They demonstrated the stability of FEM solution and its efficiency in the adaptivity strategy. 

In recent years, PINNs have emerged as a promising tool for solving PDEs by integrating physical laws directly into the neural network training process. Berardi et al., 2024
\cite{berardi2024inverse}, demonstrated the scalability and efficiency of an adaptive inverse PINN architecture for a variety of transport models, showing significant potential for complex physical systems. Moreover, Wang et al., 2023 \cite{wang2023physics}, confirmed the effectiveness of PINNs in solving variable-order space-fractional PDEs, illustrating their versatility in handling different types of PDEs beyond traditional models. Despite their potential, PINNs face challenges such as ill-conditioning when dealing with complex non-linear problems. To address these challenges, hybrid approaches have been developed that combine PINNs with traditional numerical methods.

One notable hybrid method introduced by Meetal et al., 2023 \cite{meethal2023finite}, is the FEM-enhanced Neural Network (FEM-NN) hybrid model, which integrates classical FEM with neural networks to create a robust surrogate model for both forward and inverse problems. The FEM-NN hybrid model is designed to maintain well-conditioning of the problem while ensuring accurate and physically consistent solutions. Recent studies and demonstrations show that the FEM-NN hybrid approach achieves faster convergence and greater accuracy compared to traditional PINNs, especially in complex scenarios such as steady-state convection-diffusion problems. Additionally, the integration of reduced order modeling techniques, such as Proper Orthogonal Decomposition (POD)-Galerkin by Hijazi et al., 2023 \cite{hijazi2023pod}, with PINNs has been explored to address the computational challenges of high-dimensional PDEs. By coupling POD-Galerkin reduced order models with PINNs, researchers have been able to solve inverse problems more efficiently, reducing computational costs while enhancing solution accuracy. Finally, the Galerkin-FEM (GFEM) introduced a feasible computational procedure guarantees stability conditions for a wide range of practical applications~\cite{ali2022advanced}. GFEM has shown significant promise in providing detailed insights and highly precise computational results. Its adaptability allows it to be effectively employed in various scientific and engineering fields, from fluid dynamics to structural analysis. Moreover, the method's robustness in handling complex boundary conditions makes it a go-to approach for tackling sophisticated nonlinear PDE problems.

The next section will dig deeper into the theoretical and practical background of the Galerkin method and PINNs for solving PDEs, offering a comparative analysis of their applications and effectiveness.

\section{Methodology}\label{sec3}

\subsection{Galerkin Finite Element Method}\label{subsec1}

The study focused on nonlinear PDEs are characterized by their dependence on a spatial domain \(\Omega\) and a temporal domain \([0, T]\). In the classical GFEM, we first discretize the spatial domain in $q$ number of subdomains and aim to find an approximate piecewise polynomial $\hat{u}$ for each subdomain. If $p$ is the number of nodal point per subdomain and $q$ is the number of subdomains over the whole domain, then the number of unknowns, which are time-dependant constants or variable, is given by $N=(p-1)\times q+1$. Let $\hat{u}(x,t)= \sum_{j=1}^p c_j(t)v_j(x)$ denotes the trial approximation where $v(x)$ is the quadratic shape functions. Then the weighted residual equation will be,

\begin{align} \label{gfem}
&\int_\Omega v_i(x)\left[\frac{\partial \hat{u}}{\partial t}+ \mathcal{N}[\hat{u}]  \right]dx =0 \notag, \\
&\int_\Omega v_i(x)\left[\frac{\partial \left(\sum_{j=1}^p c_j(t)v_j(x) \right) }{\partial t} + \mathcal{N}\left[\left(\sum_{j=1}^p c_j(t)v_j(x) \right) \right]  \right]dx =0 \notag, \\
&\sum_{j=1}^{p}\frac{dc_j(t)}{dt} \int_\Omega\left[v_i(x)v_j(x) \right]dx +\sum_{j=1}^{p}c_j(t) \int_\Omega \left[ v_i(x)\mathcal{N}(v_j(x))\right]dx=0.
\end{align}
Here, the choice of the function $v(x)$ and $\hat{u}$ is such that $\hat{u}, v \in H_0^1 (\Omega)$, the Sobolev space whose values are within $(0,1)$ and either $0$ or $1$ at the boundary $\partial \Omega$. This formulation results in a system of ordinary differential equations with respect to $t$, with $p$ equations for each subdomain. The coefficient matrices are of $p\times p$ size and after assembling over the whole domain the final co-efficient matrices will be  $\mathcal{R}$ and $\mathcal{K}$ of size $N\times N$ and the system of ordinary differential equations will have $N$ equation as follows,

\begin{align} \label{gfem2}
    \mathcal{R}\frac{d c(t)}{dt} + \mathcal{K}c(t) = \mathcal{F}.
\end{align}
To solve the system of equations \eqref{gfem2}, we have used the $\alpha$ family of approximation which converts the weighted average of the rate of change into a difference quotient:
\begin{align*}
    (1-\alpha)\frac{dc(t)}{dt}|_j + \alpha \frac{dc(t)}{dt}|_{j+1} \approx \frac{c(t)_{j+1}-c(t)_j}{\delta t} \qquad \text{where } 0 \le \alpha \le 1.
\end{align*}
The equation \eqref{gfem2} leads to an $N \times N$ system of nonlinear equations, which can be solved using suitable iterative methods. Finally, we obtain the solution of the system of equations for $c(t)$ and by substituting those results into the respective subdomains, we will get the piecewise polynomials for each subdomain.

\subsection{Physics-Informed Neural Network}\label{subsec2}

In the PINN method, the general form of these PDEs is represented as:
\begin{align} 
\frac{\partial u}{\partial t} + \mathcal{N}[u] = 0, \quad x \in \Omega, \quad t \in [0, T] \label{eq44},
\end{align}
where \(u(x,t)\) represents the unknown solution, \(\mathcal{N}[u]\) is a nonlinear differential operator, and \(\Omega\) is a subset of \(\mathbb{R}^D\).

The problem is supplemented with initial and boundary conditions:
\begin{align} 
u(x, 0) = h(x), \quad x \in \Omega
\end{align}
\begin{align} 
u(x, t) = g(x, t), \quad x \in \partial \Omega, \, \quad t \in [0, T].
\end{align}
Here, \(h(x)\) and \(g(x, t)\) are the initial and boundary condition functions, respectively.

PINN use a deep neural network \(u_{\theta}(x,t)\) to approximate the unknown solution \(u(x,t)\). \(\theta\) denotes the network's parameters and these parameters include weight matrices and bias vectors across the various layers of the network. These parameters determine how the network processes the input spatial and temporal coordinates to produce the predicted solution.

\begin{figure}[H] 
    \centering
    \includegraphics[width=0.8\textwidth]{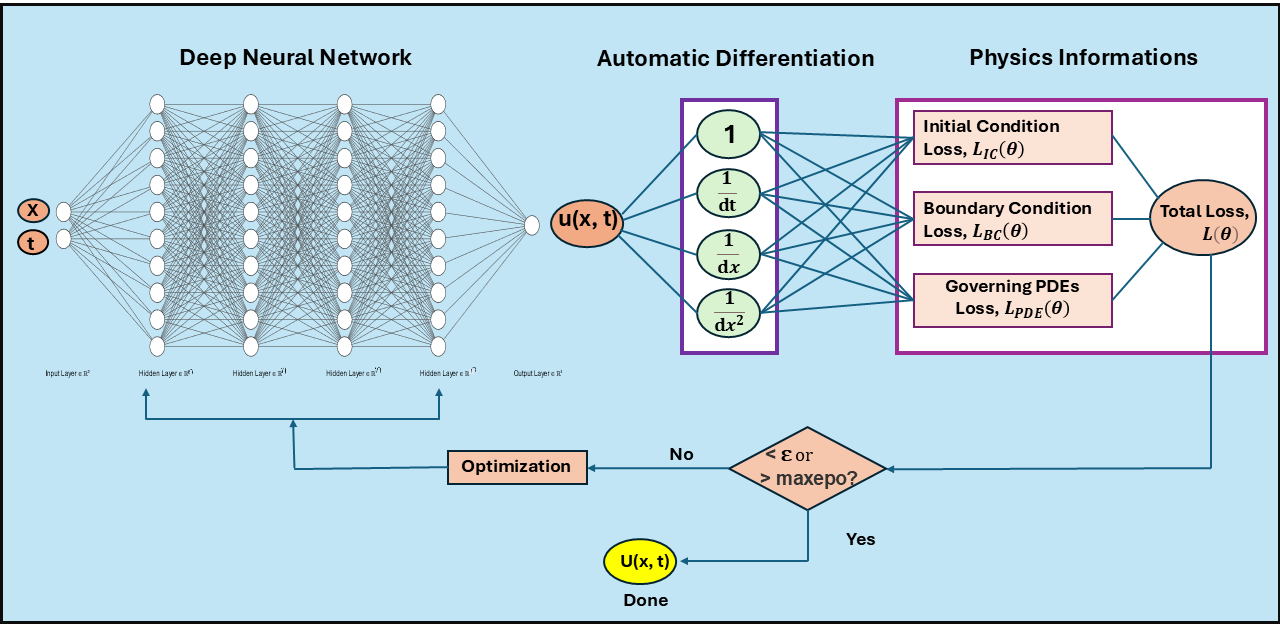}
    \caption{Neural Network Architecture for Physics-Informed Neural Network.}
    \label{fig:nn_architecture} 
\end{figure}

The residual of the PDE, denoted by \(f(t,x)\), is defined as:
\begin{align}
f(t,x) := \frac{\partial u_{\theta}}{\partial t} + \mathcal{N}[u_{\theta}].
\end{align}
In the training process of the PINN model, the objective is to minimize the discrepancy between the predicted solution and the constraints imposed by the PDE, as well as the initial and boundary conditions (Figure-\ref{fig:nn_architecture}). This is achieved by optimizing the parameters \(\theta\) through the minimization of a total loss function \(L(\theta)\), which consists of three components:
\begin{align}
L(\theta) = L_{\text{PDE}}(\theta) + L_{\text{IC}}(\theta) + L_{\text{BC}}(\theta),
\end{align}
where: 
\begin{align}
L_{\text{PDE}}(\theta) = \frac{1}{N_{\text{PDE}}} \sum_{i=1}^{N_{\text{PDE}}} \left| f(x_i, t_i, u_{\theta}, \frac{\partial u_{\theta}}{\partial t}, \frac{\partial u_{\theta}}{\partial x}, \frac{\partial^2 u_{\theta}}{\partial x^2}) \right|^2,
\end{align}
\begin{align}
L_{\text{IC}}(\theta) = \frac{1}{N_{\text{IC}}} \sum_{i=1}^{N_{\text{IC}}} \left| u_{\theta}(x_{\text{IC}}^i, t_{\text{IC}}^i) - h(x_{\text{IC}}^i) \right|^2,
\end{align}
\begin{align}
L_{\text{BC}}(\theta) = \frac{1}{N_{\text{BC}}} \sum_{i=1}^{N_{\text{BC}}} \left| u_{\theta}(x_{\text{BC}}^i, t_{\text{BC}}^i) - g(x_{\text{BC}}^i, t_{\text{BC}}^i) \right|^2.
\end{align}
Here, \(L_{\text{PDE}}(\theta)\), \(L_{\text{IC}}(\theta)\), \(L_{\text{BC}}(\theta)\) are the residual loss of PDEs, initial and boundary conditions loss respectively. The number of collocation points of the computational domain, initial, and boundary conditions are \(N_{\text{PDE}}\), \(N_{\text{IC}}\), \(N_{\text{BC}}\) respectively.

 In this method, a fully connected feedforward neural network architecture was used. The input layer of the network was designed to process the spatial and temporal coordinates \((x, t)\), while the output layer generated the approximated solution \(u_{\theta}(x, t)\). The hidden layers of the network utilized the hyperbolic tangent (tanh) activation function, which was chosen for its ability to capture complex, non-linear relationships in the underlying PDEs. The architecture was composed of four hidden layers, each containing 20 neurons. To enhance the model's generalization capabilities and mitigate the risk of overfitting, L2 regularization was integrated into the training process.

The training process involved optimizing the network parameters \(\theta\) by minimizing a loss function through the use of optimization algorithms. The training process continued either until the model completed the predefined maximum number of epochs or until the stopping criterion was met (Figure-\ref{fig:nn_architecture}). The stopping criterion was based on the difference between the total loss at the current epoch and the previous epoch; if this difference became smaller than a predefined tolerance \( \epsilon \), training was halted early.

Finally, the trained PINN model was deployed to predict the solution of the PDE at any spatial-temporal point within the domain.

The performance of the PINN model is evaluated using the absolute error metric \(\|\epsilon\|_1\), which measures the difference between the true solution \(u(x_i, t_i)\) and the predicted solution \(u_{\theta}(x_i, t_i)\) across the domain:
\begin{align}
\|\epsilon\|_1 = \frac{1}{N} \sum_{i=1}^{N} \left| u(x_i, t_i) - u_{\theta}(x_i, t_i) \right|
\end{align}
 where, N is the number of data points. The absolute error metric is well-suited for evaluating the performance of models solving nonlinear PDEs due to its simplicity, interpretability. It measures the difference between predicted and true values, giving a straightforward understanding of how well the model performs. Unlike squared error, it does not overly penalize large deviations, making it more robust to outliers. Since nonlinear PDEs describe smooth, continuous processes, absolute error offers a clear and balanced view of model accuracy across the entire domain.

Visual comparisons were also conducted between the predicted and true solutions to assess the model’s accuracy. Additionally, the performance of PINN and GFEM was evaluated through various statistical analyses for error assessment, including the RMSE, the standard deviation, the Wilcoxon Signed-Rank Test and the coefficient of variation (CV).

\section{Experimental Settings}\label{sec4}

In this study, we undertook the task of solving and conducting a comparative analysis of four significant nonlinear PDEs commonly encountered across various fields of science and engineering. The first equation we considered was the nonlinear Burgers’ Equation, which plays a crucial role in modeling a range of physical phenomena, including fluid dynamics, shock waves, traffic flow, and turbulence. This equation is often used to describe the propagation of waves and interactions in nonlinear media. Next, we explored Fisher’s Equation, which has widespread applications in heat and mass transfer, biology, and ecology. It is particularly known for modeling the spread of advantageous genetic traits within a population, making it a cornerstone in the study of population dynamics and evolutionary biology. The third equation we examined was the Burgers-Huxley Equation, a hybrid of the Burgers’ and Huxley equations, which finds utility in areas such as fluid mechanics, chloride diffusion in concrete, mathematical biology, air pollutant emissions, population genetics, nonlinear acoustics, and combustion theory. Its versatility makes it critical for understanding complex behaviors in both natural and engineered systems. Finally, we investigated the Newell-Whitehead-Segel Equation, a key model for studying the interaction between diffusion and nonlinear reaction terms. This equation is particularly valuable in fluid mechanics, biological systems, and even in financial modeling, providing insights into diffusion processes coupled with nonlinear dynamics.

In implementing these models, we employed several techniques to ensure the accuracy and robustness of our results. An important part of the experimental setup was the use of learning rates and L2 (Ridge) regularization. These elements were introduced to help the model converge and maintain stability during training. Regularization also played a pivotal role in preventing overfitting, ensuring that our models generalized well across different scenarios. In the training process, the Adam optimizer was used, with a learning rate of \(1 \times 10^{-3}\), over 20,000 epochs. During training, the combined loss function \(L(\theta)\) was minimized, and gradient clipping technique was employed to stabilize the training process. The training data included 50 points representing the initial condition, 50 points for the boundary conditions, and 10,000 collocation points, those were generated using Latin Hypercube Sampling (LHS) method.

Our experiments were carried out on an integrated Intel Iris Xe Graphics processor, which provided sufficient computational power to handle the complexity and scale of the problem. Although not a dedicated high-performance computing system, this setup proved capable of executing the intensive calculations required for both the PINNs and the GFEM approaches. For the implementation of the PINNs method, we used Python version 3.11.5, packaged by Anaconda, and integrated several essential libraries such as NumPy for numerical computations, Matplotlib for data visualization, TensorFlow for building and training the neural network models, and the `qmc` module from SciPy for generating collocation points through the Latin Hypercube Sampling method. This combination of open-source tools allowed us to efficiently approximate the solutions of these complex PDEs, leveraging the flexibility and power of Python’s ecosystem for scientific computing. For the GFEM, we turned to MATLAB, a widely trusted platform for finite element analysis and numerical simulations. MATLAB’s extensive toolsets for solving partial differential equations and its built-in support for matrix operations made it ideal for handling the GFEM implementation.

\section{Computational Results and Discussion}\label{sec5}

In this section, we solved some well-known non-linear parabolic equations, using both PINN and GFEM methods, along with their initial and boundary conditions and analytical solutions. After that, the results of the methods were presented in tabular and graphical formats.

\subsection{Problem 1:  Burgers' Equation}\label{subsec3}

The nonlinear Burgers' equation is a well-studied PDE that has found widespread applications in fluid dynamics, magnetohydrodynamics, and other areas of applied science and engineering. In this work, we consider the Burgers' equation with homogeneous Dirichlet boundary conditions \cite{wood2006exact}, which can be expressed as:
\begin{align}
    \frac{\partial u(x,t)}{\partial t} &= \frac{1}{\text{Re}} \frac{\partial^2 u(x,t)}{\partial x^2} - u(x,t) \frac{\partial u(x,t)}{\partial x}, \quad x \in \Omega \equiv [0, 1], \; t > 0, \label{eq1_burger}
\end{align}
with the boundary conditions:
\begin{align}
    u(0,t) &= u(1,t) = 0, \quad t > 0, \; x \in \partial \Omega, \label{eq2_burger}
\end{align}
and the initial condition:
\begin{align}
    u(x,0) &= \frac{2\pi \sin(\pi x)}{\text{Re}(\sigma + \cos(\pi x))}, \quad x \in \Omega, \label{eq3_burger}
\end{align}
where \(\text{Re} > 0\) is the Reynolds number and \(\sigma > 1\) is a parameter.

The analytical solution of the problem is given by:
\begin{align}
    u(x,t) &= \frac{2\pi e^{-\pi^2 t/\text{Re}} \sin(\pi x)}{\text{Re}(\sigma + e^{-\pi^2 t/\text{Re}} \cos(\pi x))}, \label{eq4_burgers}
\end{align}
where the parameter values are \(\text{Re} = 1\) and \(\sigma = 2\).

\begin{table}[H]
\centering
\caption{The accuracy of GFEM and PINN for different time points of problem 1.}
\label{tab:accuracy_comparison1}
\vspace{10pt}
\small
\begin{tabularx}{\textwidth}{c c >{\centering\arraybackslash}X >{\centering\arraybackslash}X}
\toprule
\textbf{t} & \textbf{x} & \textbf{Absolute Error (GFEM) \cite{ali2022advanced}} & \textbf{Absolute Error (PINN)} \\
\midrule
0.02 & 0.1 & \num{4.20e-03} & \num{5.09e-04} \\
     & 0.2 & \num{9.10e-03} & \num{3.37e-04} \\
     & 0.4 & \num{2.16e-02} & \num{2.42e-04} \\
     & 0.6 & \num{2.51e-02} & \num{8.70e-05} \\
     & 0.8 & \num{6.83e-03} & \num{3.90e-05} \\
     & 1.0 & \num{0.00} & \num{1.17e-03} \\
\cmidrule{1-4}
0.04 & 0.1 & \num{1.21e-02} & \num{3.87e-04} \\
     & 0.2 & \num{2.50e-02} & \num{8.00e-05} \\
     & 0.4 & \num{5.07e-02} & \num{2.92e-04} \\
     & 0.6 & \num{4.83e-02} & \num{2.63e-04} \\
     & 0.8 & \num{9.19e-03} & \num{1.62e-04} \\
     & 1.0 & \num{0.00} & \num{3.90e-05} \\
\bottomrule
\end{tabularx}
\end{table}

\begin{figure}[H]
    \centering
    \includegraphics[width=1.0\textwidth]{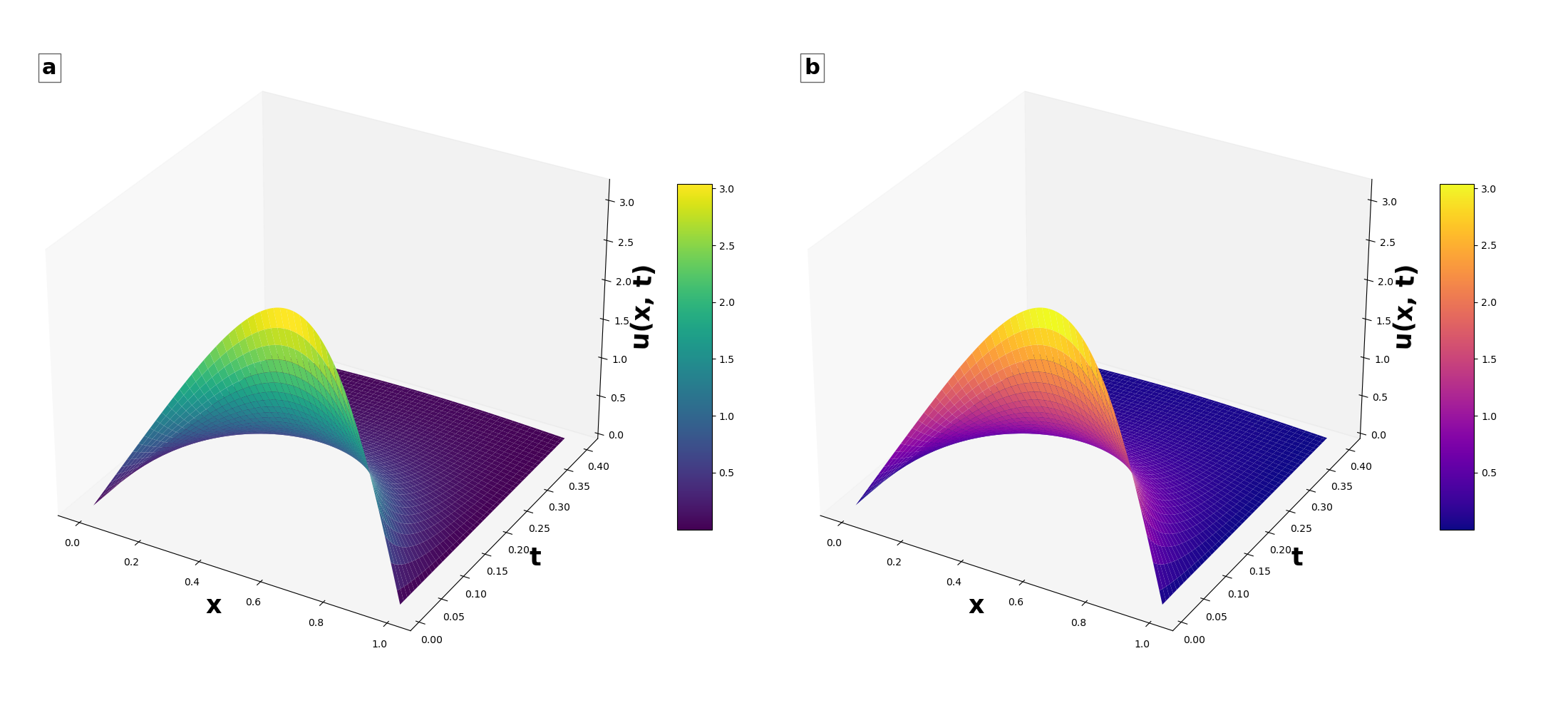}
    \caption{The 3D surface plot representations showed (a) the exact solutions and (b) the PINN solutions for Problem 1 at different time instances.}
    \label{fig:burger_solution}
\end{figure}

\begin{figure}[H]
    \centering
    \includegraphics[width=1.0\textwidth]{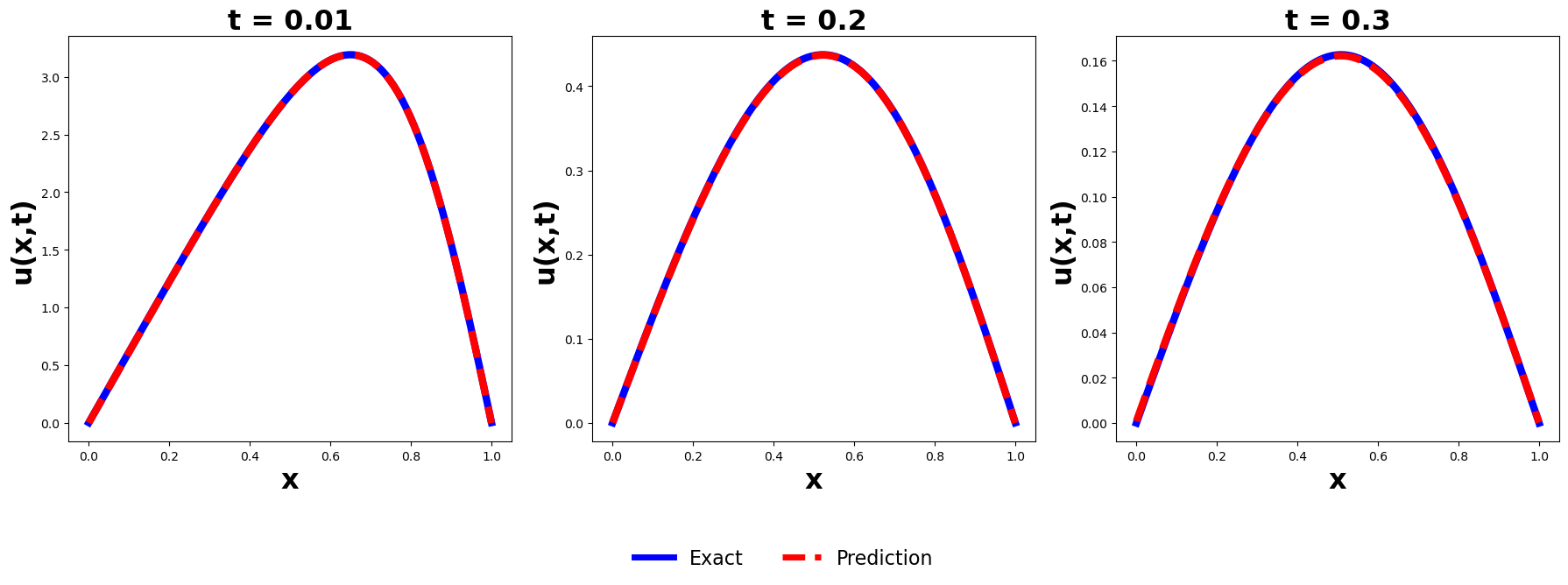}
    \caption{ Comparison of PINN solutions with exact analytical solutions for Problem 1 at time points: t=0.01, t=0.2, and t=0.3. The blue solid lines represented the exact solutions, while the red dashed lines indicated the PINN predictions.}
    \label{fig:burger_difftime}
\end{figure}

\begin{figure}[H]
\centering
\includegraphics[width=1.0\textwidth]{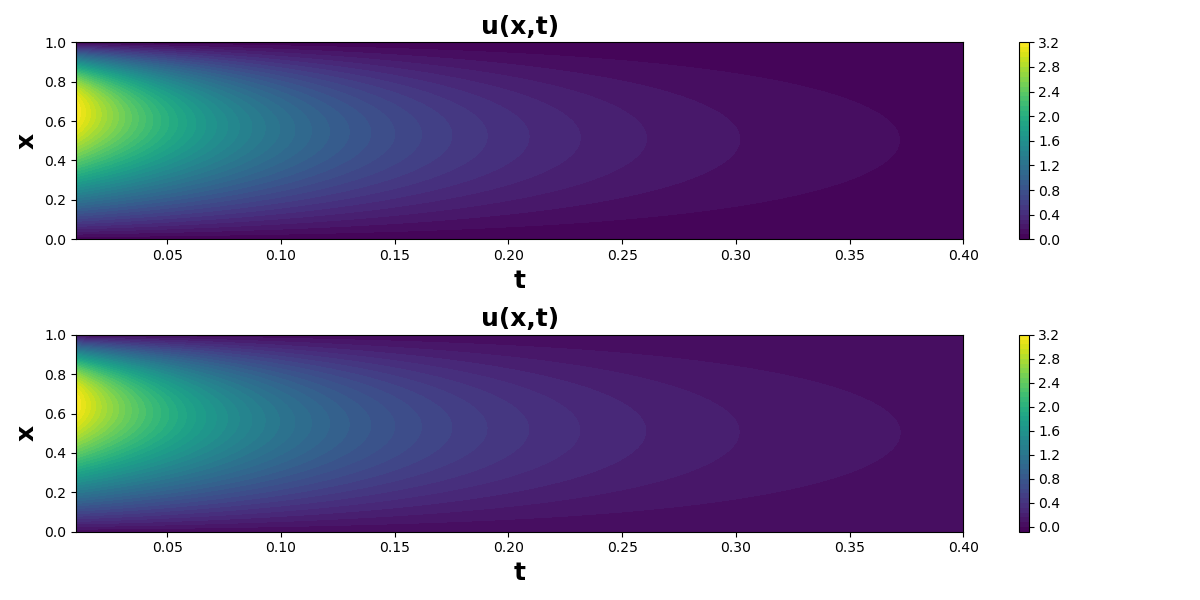}
\caption{The figure presented the exact solutions at the top and the corresponding predicted solutions obtained using the PINN for Problem 1 at various time instances in 2D distribution plot at the bottom.}
\label{fig:burger_solution_2d}
\end{figure}
The accuracy of the PINN and the GFEM of the Burgers' equation (\ref{eq1_burger} - \ref{eq3_burger}) was presented  in Table \ref{tab:accuracy_comparison1} and the Figures (\ref{fig:burger_solution} - \ref{fig:burger_solution_2d}) illustrated the analytical solution and PINN approximation. The obtained GFEM numerical results was presented in \ref{tab:accuracy_comparison1} using 40 linear elements with a time step size $k=0.01$, a spatial step size $h=0.025$, and the Crank-Nicolson scheme with second-order accuracy.

\subsection{Problem 2: Fisher’s Equation}\label{subsec4}

The Fisher’s equation with non-homogeneous Neumann boundary conditions is a fundamental model in various fields such as heat and mass transfer, biology, and ecology \cite{wazwaz2004analytic}. The equation of interest is given by:
\begin{align}
    \frac{\partial u(x, t)}{\partial t} &= \frac{\partial^2 u(x, t)}{\partial x^2} + 6u(x, t)(1 - u(x, t)), \quad x \in \Omega \equiv [0, 1], \; t > 0,
    \label{eq:fisher_equ1}
\end{align}
with the boundary conditions:
\begin{align}
    u(0, t) &= \left(1 + e^{-5t}\right)^{-2}, \quad t > 0, \, x \in \partial \Omega, \\
    u(1, t) &= \left(1 + e^{1-5t}\right)^{-2}, \quad t > 0, \, x \in \partial \Omega.
    \label{eq:fisher_equ2}
\end{align}
and the initial condition:
\begin{align}
u(x, 0) &= \left(1 + e^{x}\right)^{-2}, \quad x \in \Omega. \label{eq:fisher_equ3}
\end{align}

Here, \( u(x, t) \) represents the population density or another relevant quantity, with \( \Omega \) denoting the spatial domain \( [0, 1] \). The exact solution of the equation is:
\begin{align}
u(x, t) &= \left(1 + e^{x - 5t}\right)^{-2}. \label{eq:fisher_exact_solution}
\end{align}
This equation describes the evolution of \( u \) over time \( t \), capturing phenomena where \( u \) exhibits logistic growth or diffusion processes influenced by the interplay between diffusion and reaction terms.

\begin{table}[H]
\centering
\caption{The accuracy of GFEM and PINN for different time points of problem 2.}
\label{tab:accuracy_comparison2}
\vspace{10pt}
\small
\begin{tabularx}{\textwidth}{c c >{\centering\arraybackslash}X >{\centering\arraybackslash}X}
\toprule
\textbf{t} & \textbf{x} & \textbf{Absolute Error (GFEM) \cite{ali2022advanced}} & \textbf{Absolute Error (PINN)} \\
\midrule
0.05 & 0.1 & \num{3.02e-03} & \num{3.67e-04} \\
     & 0.2 & \num{7.91e-04} & \num{2.27e-04} \\
     & 0.4 & \num{1.35e-03} & \num{3.40e-05} \\
     & 0.6 & \num{1.51e-03} & \num{1.13e-04} \\
     & 0.8 & \num{3.87e-04} & \num{1.48e-04} \\
     & 1.0 & \num{9.67e-18} & \num{1.32e-04} \\
\cmidrule{1-4}
0.10 & 0.1 & \num{1.33e-03} & \num{4.19e-04} \\
     & 0.2 & \num{2.78e-03} & \num{3.35e-04} \\
     & 0.4 & \num{6.27e-03} & \num{8.70e-05} \\
     & 0.6 & \num{5.79e-03} & \num{2.90e-05} \\
     & 0.8 & \num{2.71e-03} & \num{1.76e-04} \\
     & 1.0 & \num{6.32e-18} & \num{2.34e-04} \\
\bottomrule
\end{tabularx}
\end{table}

\begin{figure}[H]
    \centering
    \includegraphics[width=1.0\textwidth]{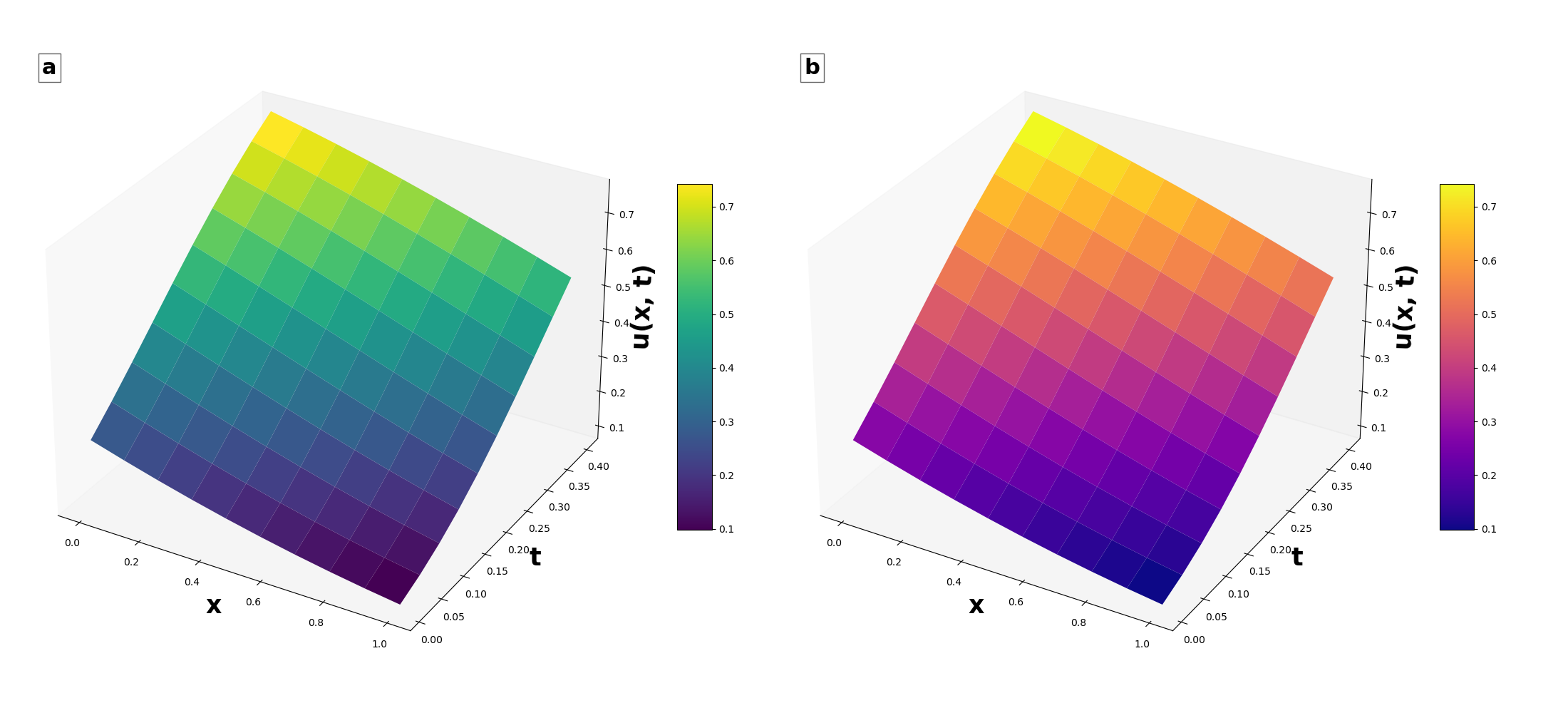}
    \caption{The 3D surface plot representations showed (a) the exact solutions and (b) the PINN solutions for Problem 2 at different time instances.}
    \label{fig:fisher_solution}
\end{figure}

\begin{figure}[H]
    \centering
    \includegraphics[width=1.0\textwidth]{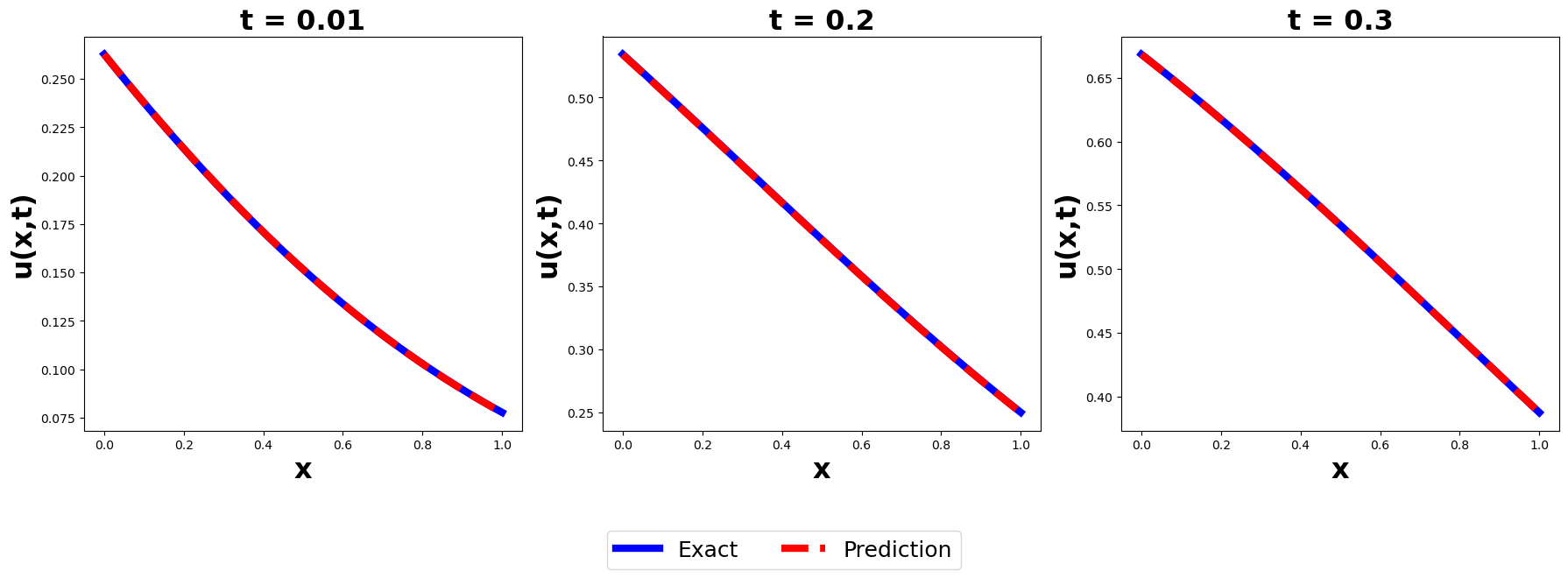}
    \caption{Comparison of the PINN predictions with the analytical solution for Problem 2 at various time points t=0.01, t=0.2, and t=0.3. The solid blue lines represented the exact analytical solutions, while the dashed red lines indicated the PINN predictions.}
    \label{fig:fisher_difftime}
\end{figure}

\begin{figure}[H]
    \centering
    \includegraphics[width=1.0\textwidth]{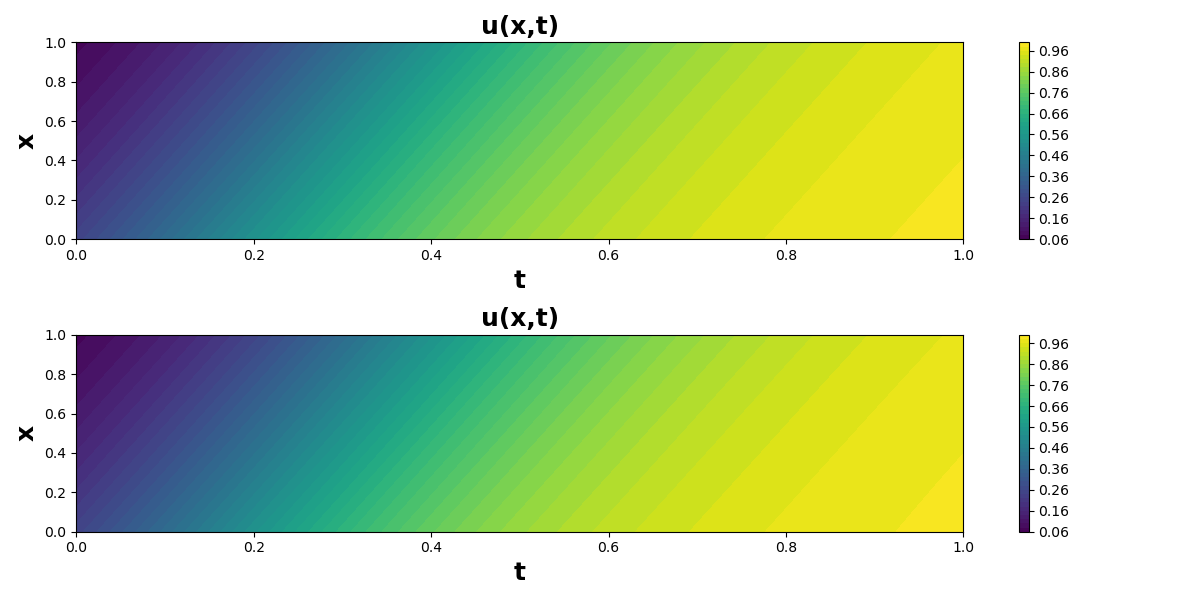}
    \caption{The figure presented the exact solutions at the top and the corresponding predicted solutions obtained using the PINN for Problem 2 at various time instances in 2D distribution plot at the bottom.}
    \label{fig:fisher_solution_2d}
\end{figure}
The accuracy of the PINN and GFEM methods of the Fisher’s equation was presented in Table \ref{tab:accuracy_comparison2}. The Figures \ref{fig:fisher_solution} - \ref{fig:fisher_solution_2d} illustrated the approximated and the analytical solution using the PINN method through a 3D surface plot, a comparison line plot at different time points, and a 2D distribution plot, respectively. For the GFEM method, we used 40 linear elements with a time step size of (k=0.01), a spatial step size of (h=0.025), and the backward difference scheme with first-order accuracy to obtain the results.

\subsection{Problem 3:  Burgers-Huxley Equation}\label{subsec5}

The Burgers-Huxley equation is a well-known nonlinear parabolic partial differential equation that has found applications in various fields, including fluid mechanics, biology, and combustion theory \cite{mittal2012cubic, zhou2011linearly}. In this work, we consider the Burgers-Huxley equation with no-flux boundary conditions, which is expressed as:
\begin{align}
    \frac{\partial u}{\partial t} &= \frac{\partial^2 u}{\partial x^2} - \phi u \frac{\partial u}{\partial x} + \epsilon u(x,t)(u(x,t) - 1)(\tau - u(x,t)), \quad x \in [-25, 17], \; t > 0, \label{eq1}
\end{align}
with the boundary conditions:
\begin{align}
    \left. \frac{\partial u}{\partial x} \right|_{x=-25} &= \left. \frac{\partial u}{\partial x} \right|_{x=17} = 0, \quad t > 0, \; x \in \partial \Omega, \label{eq2}
\end{align}
and the initial condition:
\begin{align}
    u(x, 0) &= \frac{3}{2} + \frac{1}{2} \tanh\left(\frac{x}{2}\right), \quad x \in \Omega. \label{eq3}
\end{align}
Here, \(\phi\), \(\epsilon\), and \(\tau\) are real parameters. 

The exact solution of this equation is given by,
\begin{align}
    u(x, t) &= \frac{3}{2} + \frac{1}{2} \left[\tanh\frac{1}{2}\left(x + 3t\right)\right], \label{eq4}
\end{align}
where the parameter values are \(\phi = -1\), \(\epsilon = 1\), and \(\tau = 2\).
\begin{table}[H]
\centering
\caption{The accuracy of GFEM and PINN for different time levels of problem 03.}
\label{tab:accuracy_comparison3}
\vspace{10pt}
\small
\begin{tabularx}{\textwidth}{c c >{\centering\arraybackslash}X >{\centering\arraybackslash}X}
\toprule
\textbf{t} & \textbf{x} & \textbf{Absolute Error (GFEM) \cite{ali2022advanced}} & \textbf{Absolute Error (PINN)} \\
\midrule
0.10 & -25.00 & \num{4.73e-13} & \num{4.27e-03} \\
     & -20.80 & \num{1.86e-10} & \num{7.27e-04} \\
     & -12.40 & \num{8.26e-07} & \num{2.53e-03} \\
     & -4.00 & \num{3.49e-03} & \num{5.37e-03} \\
     & 0.20 & \num{3.50e-03} & \num{6.89e-04} \\
     & 4.40 & \num{1.22e-03} & \num{5.41e-03} \\
      & 12.80 & \num{2.79e-07} & \num{3.04e-03} \\
     & 17.00 & \num{8.48e-09} & \num{6.30e-03} \\
\cmidrule{1-4}
0.20 & -25.00 & \num{3.84e-12} & \num{5.38e-03} \\
     & -20.80 & \num{4.09e-011} & \num{1.00e-03} \\
     & -12.40 & \num{1.82e-07} & \num{3.47e-03} \\
     & -4.00 & \num{7.18e-04} & \num{6.09e-03} \\
     & 0.20 & \num{2.67e-03} & \num{1.81e-03} \\
     & 4.40 & \num{7.54e-04} & \num{4.19e-03} \\
      & 12.80 & \num{1.78e-07} & \num{2.08e-03} \\
     & 17.00 & \num{4.11e-09} & \num{5.22e-03} \\
\bottomrule
\end{tabularx}
\end{table}

\begin{figure}[H] 
    \centering
    \includegraphics[width=1.0\textwidth]{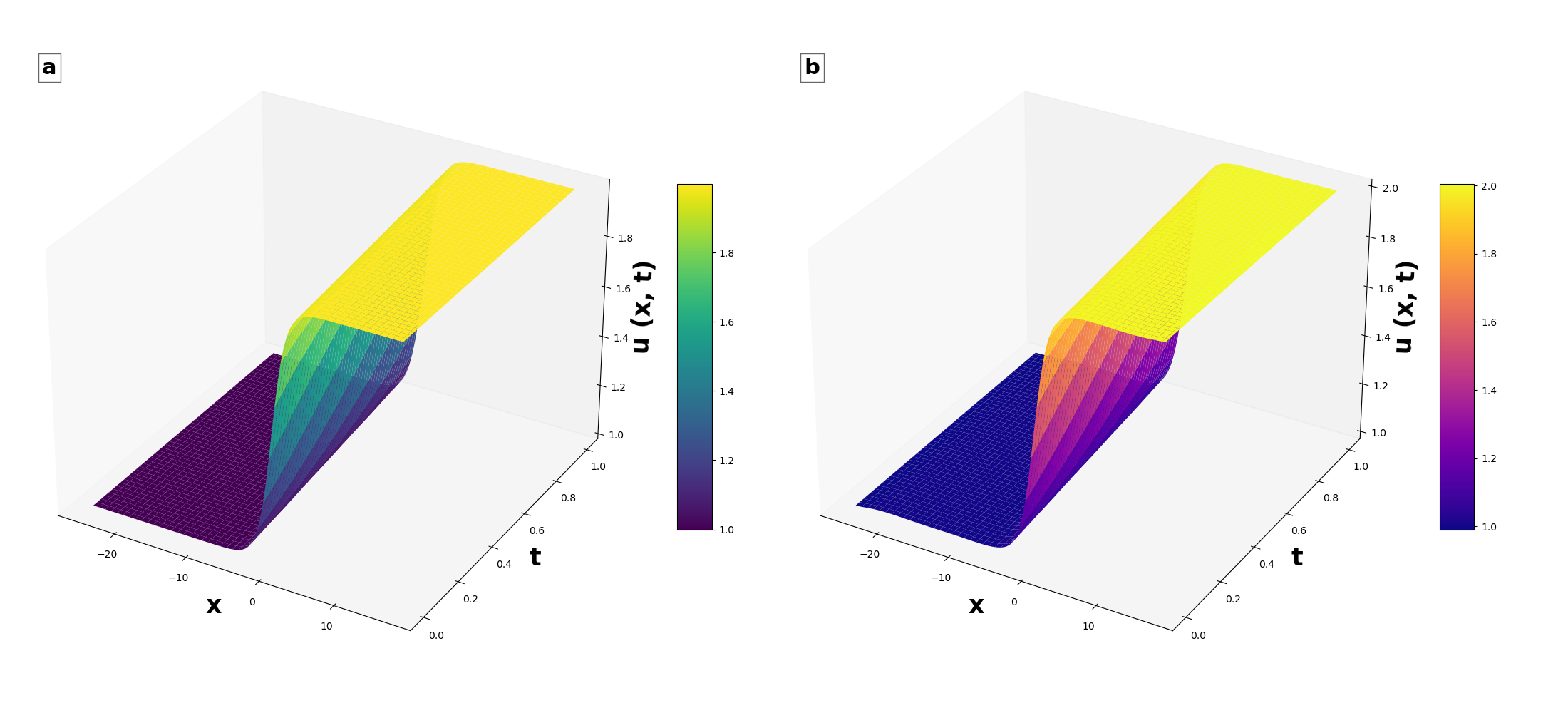}
    \caption{The 3D representations showed (a) the exact solutions and (b) the PINN solutions for Problem 2 at different time instances.}
    \label{fig:burgerhux_solution}
\end{figure}

\begin{figure}[H]
    \centering
    \includegraphics[width=1.0\textwidth]{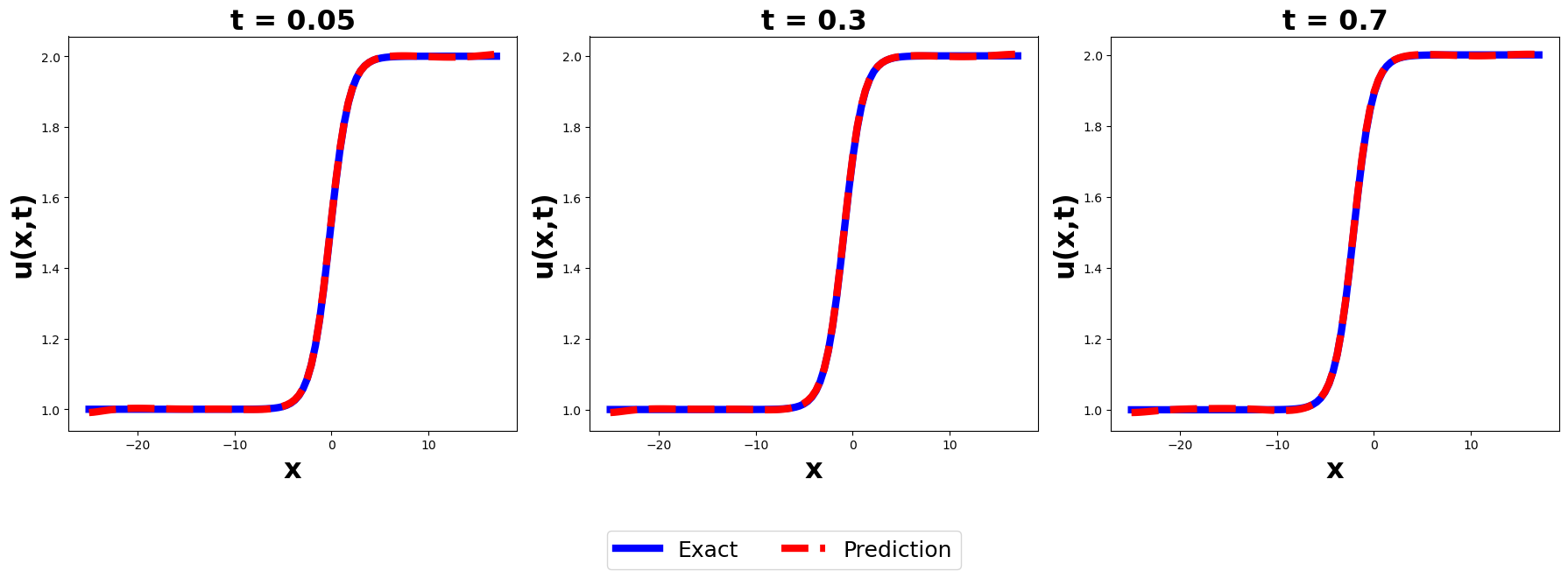}
    \caption{Depiction of the PINN performance in predicting the exact solutions for Problem 3 at distinct times: t=0.01, t=0.2, and t=0.3. The exact analytical solutions were  shown as blue solid lines, while the PINN predictions were presented as red dashed lines.}
    \label{fig:burgerhux_difftime}
\end{figure}
\begin{figure}[H]
    \centering
    \includegraphics[width=1.0\textwidth]{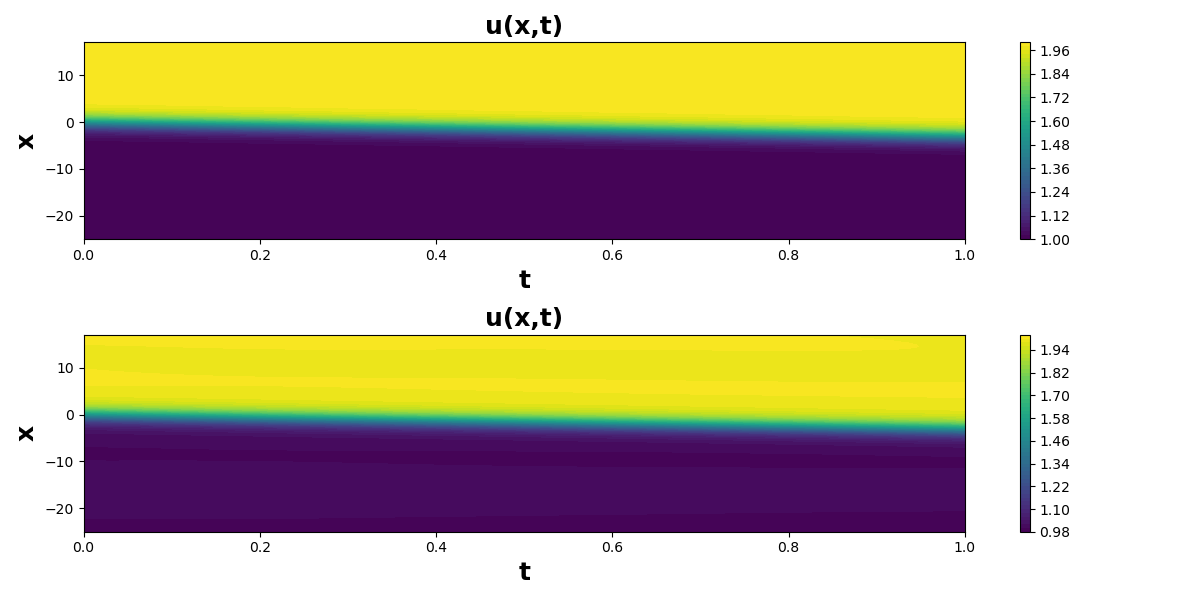}
    \caption{The figure presented the exact solutions at the top and the corresponding predicted solutions obtained using the PINN for Problem 3 at various time instances in 2D at the bottom.}
    \label{fig:burgerhux_solution_2d}
\end{figure}
The accuracy results of PINN and GFEM approaches of problem 3 were presented in Table \ref{tab:accuracy_comparison3}. The approximated and analytical solutions obtained using the PINN method were visualized in Figures \ref{fig:burgerhux_solution} - \ref{fig:burgerhux_solution_2d}, which include a 3D plot, a comparison plot at various time points, and a 2D plot.  We used 40 linear elements with a time step size k=0.01, a spatial step size h= 1.05, and the backward difference scheme with first-order accuracy.

\subsection{Problem 4: Newell-Whitehead-Segel Equation}\label{subsec6}

The Newell-Whitehead-Segel equation models the interaction between the diffusion term and the nonlinear effect of the reaction term \cite{nourazar2015exact}. This equation is given by:
\begin{align}
    \frac{\partial u(x, t)}{\partial t} &= \eta \frac{\partial^2 u(x, t)}{\partial x^2} + \mu u(x, t) - \nu u(x, t)^\rho, \quad x \in \Omega \equiv [0, 1], \; t > 0,
\end{align}
with the boundary conditions:
\begin{align}
    \left. \frac{\partial u}{\partial x} \right|_{x=0} &= -\frac{2}{\sqrt{6}}\frac{e^{-5t/6}}{\left(1 + e^{-5t/6}\right)^3}, \quad t > 0, \; x \in \partial \Omega, \\
    \left. \frac{\partial u}{\partial x} \right|_{x=1} &= -\frac{2}{\sqrt{6}}\frac{e^{(\sqrt{6} - 5t)/6}}{\left(1 + e^{(\sqrt{6} - 5t)/6}\right)^3}, \quad t > 0, \; x \in \partial \Omega,
\end{align}
and the initial condition:
\begin{align}
    u(x, 0) &= \left(1 + e^{x/ \sqrt{6}}\right)^{-2}, \quad x \in \Omega,
\end{align}
where \(\eta\), \(\mu\), and \(\nu\) are real numbers with \(\eta > 0\) and \(\rho\) is a positive integer. To approximate the solution of the Newell-Whitehead-Segel equation, we choose \(\eta = 1\), \(\mu = 1\), \(\nu = 1\), and \(\rho = 2\).

The exact solution of the equation is:
\begin{align}
    u(x, t) &= \left(1 + e^{\frac{x}{\sqrt{6}} - \frac{5t}{6}}\right)^{-2}.
\end{align}

\begin{table}[H]
\centering
\caption{The accuracy of GFEM and PINN for different time levels of problem 4.}
\label{tab:accuracy_comparison4}
\vspace{10pt}
\small
\begin{tabularx}{\textwidth}{c c >{\centering\arraybackslash}X >{\centering\arraybackslash}X}
\toprule
\textbf{t} & \textbf{x} & \textbf{Absolute Error (GFEM) \cite{ali2022advanced}} & \textbf{Absolute Error (PINN)} \\
\midrule
0.03 & 0.1 & \num{1.48e-06} & \num{9.10e-06} \\
     & 0.2 & \num{1.88e-05} & \num{1.77e-05} \\
     & 0.4 & \num{3.57e-05} & \num{3.15e-05} \\
     & 0.6 & \num{4.34e-05} & \num{3.24e-05} \\
     & 0.8 & \num{6.45e-05} & \num{2.47e-05} \\
     & 1.0 & \num{1.32e-04} & \num{1.03e-05} \\
\cmidrule{1-4}
0.05 & 0.1 & \num{3.06e-05} & \num{2.55e-05} \\
     & 0.2 & \num{5.45e-05} & \num{3.18e-05} \\
     & 0.4 & \num{7.97e-05} & \num{3.78e-05} \\
     & 0.6 & \num{9.55e-05} & \num{3.50e-05} \\
     & 0.8 & \num{1.28e-04} & \num{2.31e-05} \\
     & 1.0 & \num{2.03e-04} & \num{1.38e-05} \\
\bottomrule
\end{tabularx}
\end{table}

\begin{figure}[H]
    \centering
    \includegraphics[width=1.0\textwidth]{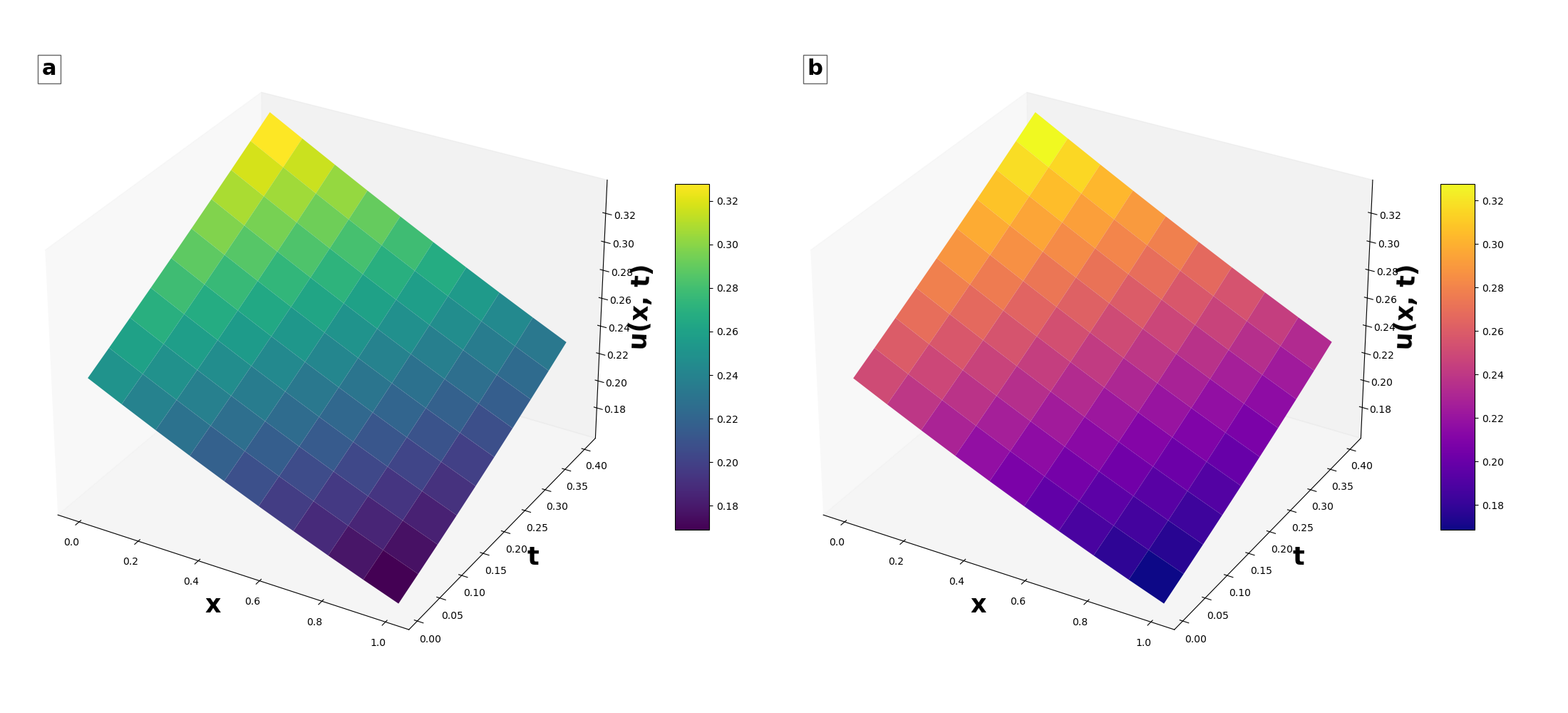}
    \caption{The 3D surface plot representations showed (a) the exact solutions and (b) the PINN solutions for Problem 2 at different time instances.}
    \label{fig:problem4_solution}
\end{figure}

\begin{figure}[H]
    \centering
    \includegraphics[width=1.0\textwidth]{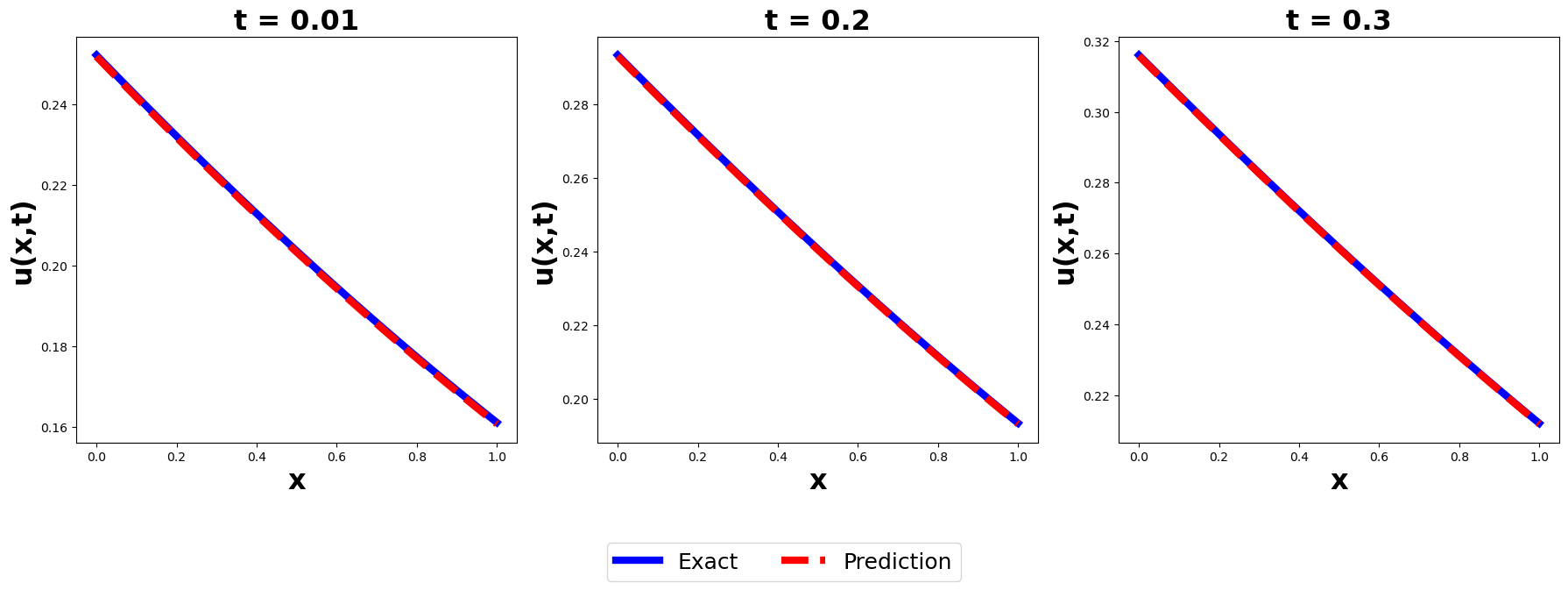}
    \caption{Comparative analysis of the PINN solutions against the exact analytical solutions for Problem 4 at distinct time points: t=0.01, t=0.2, and t=0.3. The exact solutions were depicted by solid blue lines, while the PINN predictions were shown as dashed red lines.}
    \label{fig:problem4_difftime}
\end{figure}

\begin{figure}[H]
    \centering
    \includegraphics[width=1.0\textwidth]{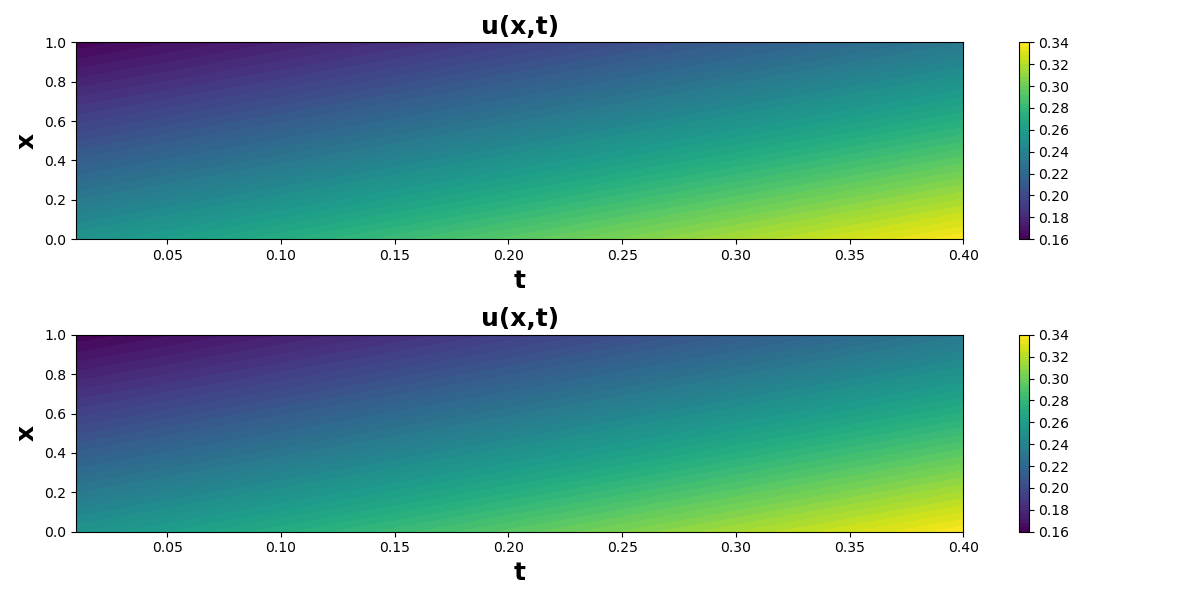}
    \caption{The figure presents the exact solutions at the top and the corresponding predicted solutions obtained using the PINN for Problem 4 at various time instances in 2D distribution plot at the bottom.}
    \label{fig:newell_solution_2d}
\end{figure}

For the Newell-Whitehead-Segel equation, the results in Table \ref{tab:accuracy_comparison4} showed the accuracy of the PINN and GFEM methods. Figures \ref{fig:problem4_solution} - \ref{fig:newell_solution_2d} illustrated the approximated and the analytical solution using the PINN method through a 3D surface plot, a comparison line plot at different time points, and a 2D distribution plot, respectively. We used 40 linear elements for GFEM with a time step size of k = 0.01 and a spatial step size of h=0.025, and the backward difference scheme with first-order accuracy. Some points of the GFEM results showed lower accuracy compared to the PINN method.

\begin{table}[H]
\centering
\caption{Comparative Performance of PINN and GFEM across Nonlinear Convection-Reaction-Diffusion Problems}
\vspace{10pt}
\label{tab:statistical_comparison}
\begin{tabularx}{\textwidth}{l c c c c c}
\toprule
\textbf{Problem} & \textbf{Method} & \textbf{RMSE} & \textbf{SDE*} & \makecell{\textbf{CV}} & \makecell{\textbf{WSRT*} \\ \textbf{(p-value)}} \\
\midrule
Burgers’ Eq.       & GFEM & 0.0177   & 0.0165  & 93.15\% & 0.0024 \\
                        & PINN & 0.0003  & 0.0003  & 99.03\%  & \\
\midrule
Fisher’s Eq.       & GFEM & 0.0022   & 0.0020 & 92.15\% & 0.0024 \\
                        & PINN & 0.0002  & 0.0001  & 63.90\%   &  \\
\midrule
Burgers-Huxley  & GFEM & 0.0008   & 0.0012 & 160.58\% & 0.0010 \\
Eq.                        & PINN & 0.0036   & 0.0019  & 52.44\%    &       \\
\midrule
Newell-Whitehead & GFEM & 0.00007 & 0.00006 & 74.58\% & 0.0034 \\
-Segel Eq.           & PINN & 0.00002 & 0.00001 & 38.43\% &       \\
\bottomrule
\end{tabularx}
\begin{tabularx}{\textwidth}{X}
\scriptsize * SDE = Standard Deviation of Error, CV = Coefficient of Variation, WSRT = Wilcoxon Signed-Rank Test \\
\end{tabularx}
\end{table}

Table 5 presents a comparative analysis of the errors (Table 1 - Table 4) in CRD problems for both the PINN and GFEM methods. For the Burgers’ equation and Fisher’s equation, PINN achieves significantly lower RMSE values (0.0004 and 0.0002, respectively) compared to GFEM (0.024 and 0.003), along with smaller standard deviations, indicating higher accuracy and greater consistency. These findings are further validated by the Wilcoxon Signed-Rank Test, which indicates statistically significant p-values supporting the performance of PINN. In contrary, for the Burgers-Huxley equation, GFEM has a lower RMSE (0.001 compared to 0.004 for PINN) and standard deviation 0.0012, suggesting slightly better accuracy than that of PINN. However, the Coeffecient of variation (CV) test reveals an important nuance in the performance that error values of GFEM fluctuate more significantly over time compared to PINN. This result indicates that while GFEM may show lower RMSE and standard deviation overall, its performance is less consistent across different time points, a critical consideration when evaluating reliability. In the Newell-Whitehead-Segel equation, PINN again outperforms GFEM, with the lowest RMSE (0.00003) and a significantly reduced standard deviation, confirming its consistency and alignment with analytical solutions.

The comparative evaluation of these metrics and the figures clearly illustrate that PINN shows consistently better performance than GFEM in nonlinear CRD problems.

The challenge of solving non-linear PDEs has led to the development of various methods, each with its strengths and challenges. The limitations of this study focused on solving one-dimensional nonlinear PDEs; multi-dimensional PDEs were not tested here. The computational cost, particularly usage of GPUs and processing time, was not thoroughly examined.

\section{Conclusion}\label{sec6}

In this study, we conducted a comprehensive comparative evaluation of the GFEM and PINN for solving four significant nonlinear CRD partial differential equations: Burgers’ Equation, Fisher’s Equation, Burgers-Huxley Equation, and the Newell-Whitehead-Segel Equation. Our findings indicate that while both methods are capable of approximating the solutions to these complex PDEs with a reasonable degree of accuracy, the PINN approach consistently outperforms GFEM in terms of accuracy and consistency across all tested problems. The PINN method demonstrated significantly better performance in accuracy and consistency over the GFEM method. Visual representations also showed that the PINN solutions closely align with the exact analytical solutions. Future research could extend the application of PINN and GFEM to solve multi-dimensional nonlinear PDEs.



\bibliographystyle{elsarticle-num} 
\bibliography{references}

\end{document}